\documentclass[prl,twocolumn,longbibliography]{revtex4-1}
\usepackage{amsmath}
\usepackage{amssymb}
\usepackage{natbib}
\usepackage{subfigure}
\usepackage{mathtools}
\usepackage{graphicx}
\usepackage{color}
\usepackage{setspace}
\usepackage{mathrsfs}
\usepackage[usenames,dvipsnames]{xcolor}
\newcommand\zi{\mathrm{i}}

\def\bdot{\raise.2em\hbox to .15em{.}}

\definecolor{gray}{gray}{0.5}

\def\bdotblack{\raise.25em\hbox to .15em{.}}

\definecolor{pinegreen}{rgb}{0.0, 0.47, 0.44}
\definecolor{debianred}{rgb}{0.84, 0.04, 0.33}

\begin{document}
\title{Exact travelling wave solutions in viscoelastic channel flow}

\author{Jacob Page$^{\dag}$} 
\author{Yves Dubief$^*$}
\author{Rich R. Kerswell$^{\dag}$}
\affiliation{$^\dag$DAMTP, Centre for Mathematical Sciences, University of Cambridge, Cambridge, CB3 0WA, UK.}
\affiliation{$^*$School of Engineering, University of Vermont, VT 05405}

\date{\today}

\begin{abstract}
    Elasto-inertial turbulence (EIT) is a new, two-dimensional chaotic flow state observed in polymer solutions with possible connections to inertialess elastic turbulence and drag-reduced Newtonian turbulence. In this Letter, we argue that the origins of  EIT are fundamentally  different from Newtonian turbulence by finding a dynamical connection  between EIT and an elasto-inertial  linear instability recently found at high Weissenberg numbers (Garg et al. {\em Phys. Rev. Lett.} {\bf 121}, 024502, 2018). This link is established by isolating the first known exact coherent structures in  viscoelastic parallel flows - nonlinear elasto-inertial travelling waves (TWs) - borne at  the linear instability and tracking them down to substantially lower Weissenberg numbers  where  EIT exists.  These TWs have a distinctive ``arrowhead' structure in the polymer stretch field and can be clearly recognised, albeit transiently, in EIT, as well as  being attractors for EIT dynamics if the  Weissenberg number is sufficiently large.  Our findings suggest that the dynamical systems picture in which Newtonian turbulence is built around the co-existence of many (unstable) simple invariant solutions populating phase space carries over to EIT, though these solutions rely on elasticity to exist.
\end{abstract}

\maketitle


The addition of only a few parts per million of long chain polymer molecules to a Newtonian solvent can fundamentally alter classical (Newtonian) turbulence at high Reynolds number ($Re \gg 1$,\,\cite{White2008}) and seed new, visually striking chaotic motion in viscosity-dominated flows, which persist even in the inertialess limit ($Re \ll 1$) -- so-called {\em elastic turbulence} \cite{Groisman2000}.  The disruption of near-wall Newtonian turbulence is well known due to the accompanied reduction in skin-friction drag (up to 80\%) and is exploited in oil pumping, for example in the trans-Alaska pipeline.  For the fluid's elasticity to manifest, the Weissenberg number ($Wi$, the ratio between a polymer relaxation timescale and a flow timescale) must be large enough to allow the polymers to stretch as they are sheared, creating an elastic tension in the streamlines. In drag-reduced flows, this effect seems to reduce the sweeps of high-momentum fluid towards the wall, though the exact mechanisms of polymeric drag reduction, and the universality of its maximum drag reduction (MDR) at 80\% \cite{Virk1970,graham2014} remain open research questions.

Recently, experiments and simulations have revealed the existence of a new turbulent flow state observed at modest inertia and elasticity ($Re=O(1000)$, $Wi=O(10)$, \cite{Samanta2013,Dubief2013}). This \emph{elasto-inertial turbulence} (EIT) is dominated by spanwise-coherent sheets in which the polymer becomes highly stretched, and attached to the sheets are regions of intense rotational and extensional flow \cite{Dubief2013,Terrapon2014}. Recent numerical simulations have confirmed that EIT is a two-dimensional phenomenon \cite{Sid2018},  while experiments in pipe flow indicate that MDR may be a feature of EIT and not a polymeric perturbation of Newtonian turbulence \cite{Choueiri2018} as has been assumed \cite{Xi2012}. Very recently, numerical simulations of EIT  have revealed the existence of a recurring coherent structure in the  turbulence - an `arrowhead' of polymer stretch - upon which EIT collapses as the Weissenberg number is increased \cite{Dubief2020}. The potential importance of EIT in drag reduced flows -- and also a possible link to elastic turbulence at $Re=0$ -- raises the  important question as to its origin. 
One candidate is a newly-discovered elasto-inertial instability \cite{Garg2018}, found in planar channel flow and pipe flow, but which exists at much higher $Wi$ than those at which EIT has been observed. The linearly unstable eigenfunctions of the instability also bear little resemblence to the recently found `arrowhead' state seen in EIT making any link unclear. 

The purpose of this Letter is to establish this link by demonstrating that the elasto-inertial travelling waves which originate at the bifurcation point found by Garg et al. \cite{Garg2018} correspond to the arrowhead coherent structures found in EIT \cite{Dubief2020}. Specifically, we show that: a) this bifurcation is substantially subcritical in $Wi$ so states connected with the instability exist at  much lower $Wi$ where EIT exists; and b) it is the upper branch of travelling waves which correspond to the arrowhead solutions not the far weaker lower branch states which resemble the eigenfunctions. Beyond the significance of isolating exact nonlinear structures in viscoelastic channel flows for the first time, our findings suggest (as conjectured by Garg et al. \cite{Garg2018}) that EIT is built around the nonlinear states which originate at an elasto-inertial instability in a similar manner to Newtonian turbulence, albeit with a completely different bifurcation structure of underlying elasto-inertial states.

%
%
\begin{figure}
    \centering
    \includegraphics[width=0.48\textwidth]{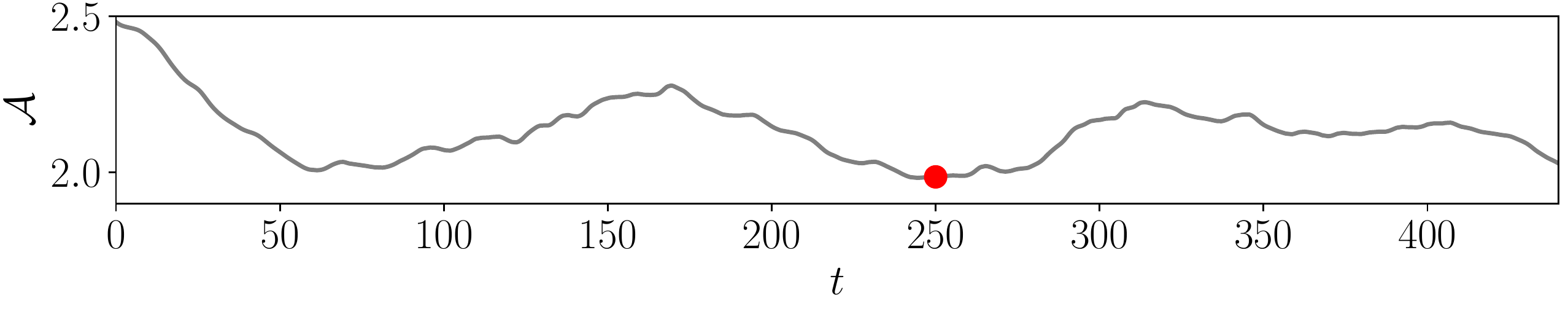}
    \includegraphics[width=0.48\textwidth]{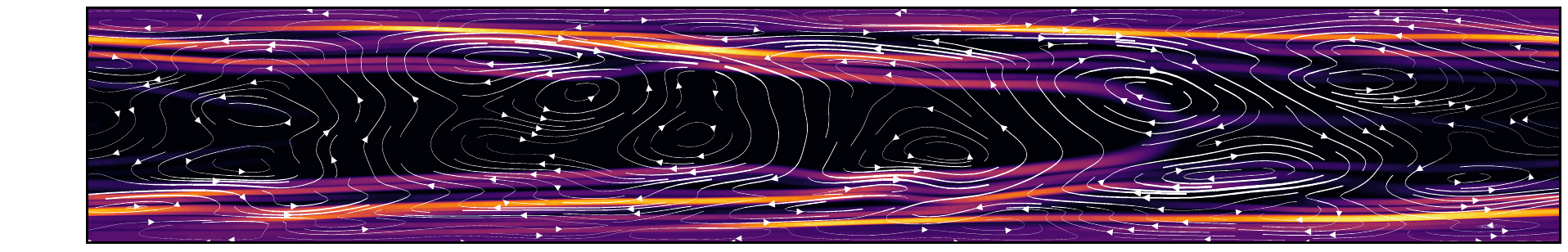}
    \includegraphics[width=0.48\textwidth]{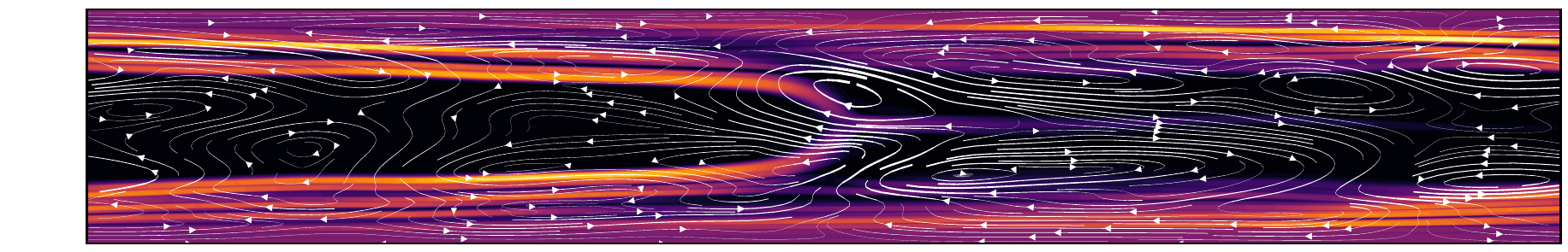}
    \caption{
        (Top) Time evolution of volume-averaged trace of the polymer conformation relative to the laminar 
        value, $\mathcal A:= \langle\text{tr}\mathbf C\rangle_V / \langle \text{tr}\mathbf C_{\text{lam}}\rangle_V$, in 
        a computation of EIT in a long domain, $l_x=4\pi$, at $Re=1000$, $Wi=20$.
        (Middle) Snapshot of the flow extracted at $t=250$ (see marker in the top panel),
    contours show $\text{tr}\mathbf C/L^2$, lines are the perturbation streamfunction for ($\mathbf u - \mathbf U_{\text{lam}}$).
        (Bottom) Snapshot of \emph{transient} EIT at $Wi=30$, with all other parameters held fixed. The flow here eventually settles
        onto a stable travelling wave (an arrowhead).}
    \label{fig:long}
\end{figure}
%
%
%

Direct numerical simulations (DNS) are performed in a 2D channel under conditions of constant mass-flux using the FENE-P model,
\begin{subequations}
\begin{align}
    \partial_t\mathbf u + \mathbf u\cdot\boldsymbol\nabla\mathbf u +\boldsymbol\nabla p &=  \frac{\beta}{Re}\nabla^2 \mathbf u 
    + \frac{(1-\beta)}{Re}\boldsymbol\nabla \cdot \mathbf T, \\
    \boldsymbol\nabla \cdot \mathbf u &= 0, \\
    \partial_t\mathbf C + \mathbf u\cdot\boldsymbol\nabla\mathbf C + \mathbf T &= \mathbf C\cdot\boldsymbol\nabla \mathbf u
    + (\boldsymbol\nabla \mathbf u)^T\cdot \mathbf C
\end{align}
where the polymeric stress, $\mathbf T$, is related to the polymer conformation tensor, $\mathbf C$, via the Peterlin function
\begin{equation}
    \mathbf T = \frac{1}{Wi}\left(\frac{\mathbf C}{1-\text{tr}\mathbf C/L^2} - \mathbf I\right).
\end{equation}
    \label{eqn:governing}
\end{subequations}
The equations are non-dimensionalised by the channel half height, $h$, and bulk velocity $U_b$, so the Reynolds and Weissenberg numbers are defined as $Re:=hU_b/\nu$ and $Wi:=\tau U_b/h$ with $\tau$ the polymer relaxation time.
The ratio of solvent to total viscosities, $\beta:=\nu_s/\nu$, is fixed at $\beta =0.9$ and the maximum extension of the polymer chains 
relative to their equilibrium length is held at $L=500$. The numerical method uses second-order finite differences in both directions which ensures the discrete conservation of mass, momentum and kinetic energy, has been extensively validated and described in detail in Sid et al. (2018).

A Newton-Krylov solver is wrapped around the DNS code to converge travelling waves (TWs) as exact solutions of the governing equations. A global diffusion term $\nabla^2 \mathbf C/ ReSc$ is added to the right hand side of (\ref{eqn:governing}c) with a Schmidt number of $Sc=10^3$ as in Sid et al. (2018). The presence of this global diffusion dramatically improves convergence properties in the Newton solver, and we obtain qualitatively similar results when timestepping the TWs with this term removed (i.e. $Sc=\infty$).
%
%
Computation and continuation of travelling waves are performed in a box of streamwise length $l_x=\pi$ at a usual resolution $N_x=128$, $N_z=513$ (others were used to check robustness) with calculations in longer boxes done with correspondingly higher stereamwise resolution to retain the same grid spacing $\Delta x$.

The time evolution of the volume-averaged trace of $\mathbf C$ in a long-box ($l_x=4\pi$) calculation at 
$(Re,Wi)=(1000,20)$ is reported in figure \ref{fig:long}, alongside a representative snapshot of the flow.
The attractor in this configuration is chaotic, and the flow shows features common to earlier computations of EIT, including the arrangement of strong regions of $\text{tr}\mathbf C$ in thin sheets which orient and stretch in the direction of the driving flow. 
Notable in the snapshot is the presence of the large ``arrowhead'' structure (about 3/4 along the channel) , which is roughly symmetric
about the channel centreline and consists of a pair of sheets which reach down into the near wall regions but also curve up to meet at $z=0$. 
As shown in the lower panel of figure \ref{fig:long}, the arrowhead becomes more pronounced with increasing $Wi$ (it is a stable attractor 
at $Wi=30$).
The emergence and stabilisation of arrowheads with increasing $Wi$ has been examined recently in \cite{Dubief2020};
they appear to be fundamental structures underpinning EIT. We now show how arrowheads connect to the centre mode instability discovered in \cite{Garg2018}.

%
%
\begin{figure*}
    \centering
    \includegraphics[width=0.56\textwidth]{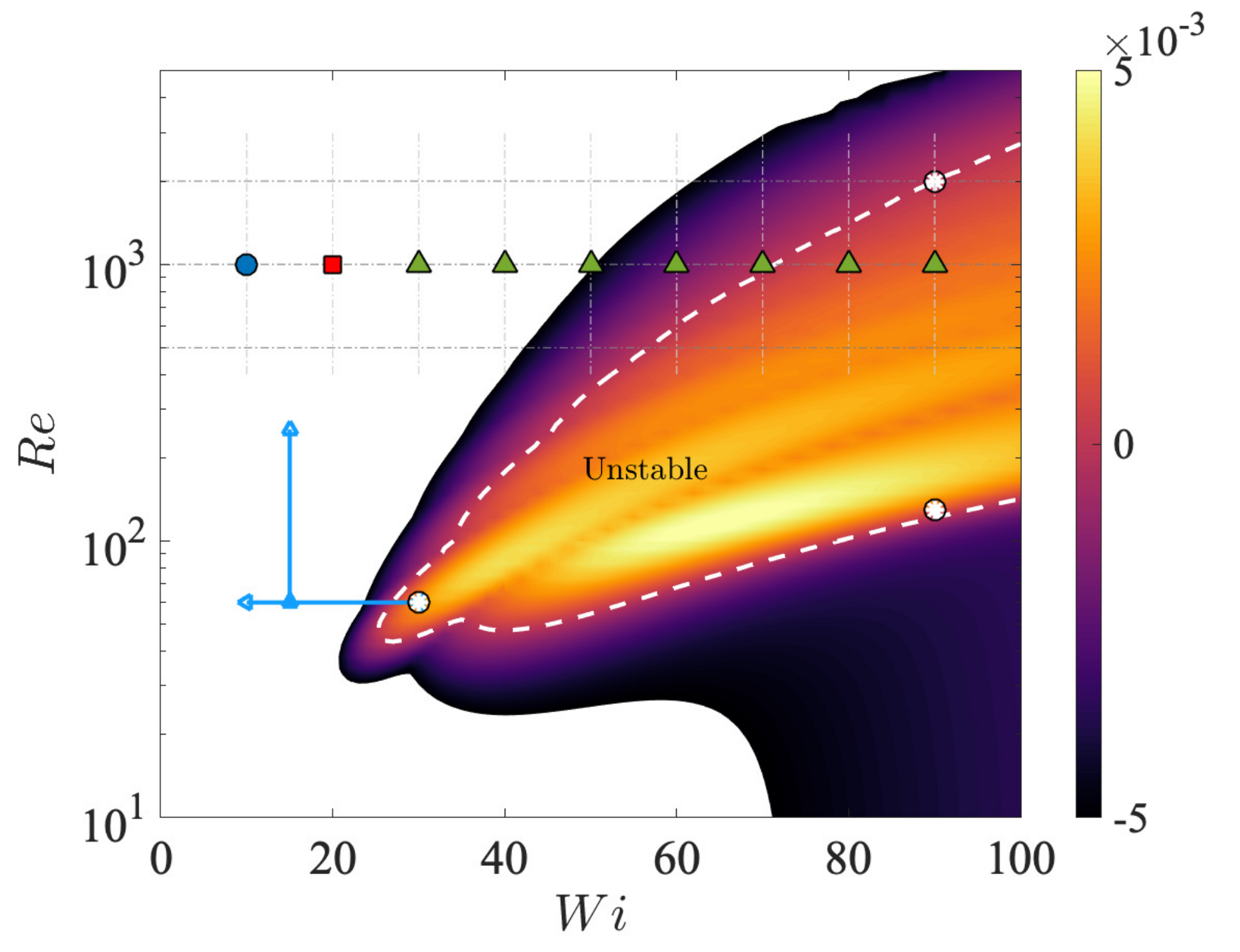}
    \includegraphics[width=0.32\textwidth, trim=0 -12 0 0, clip]{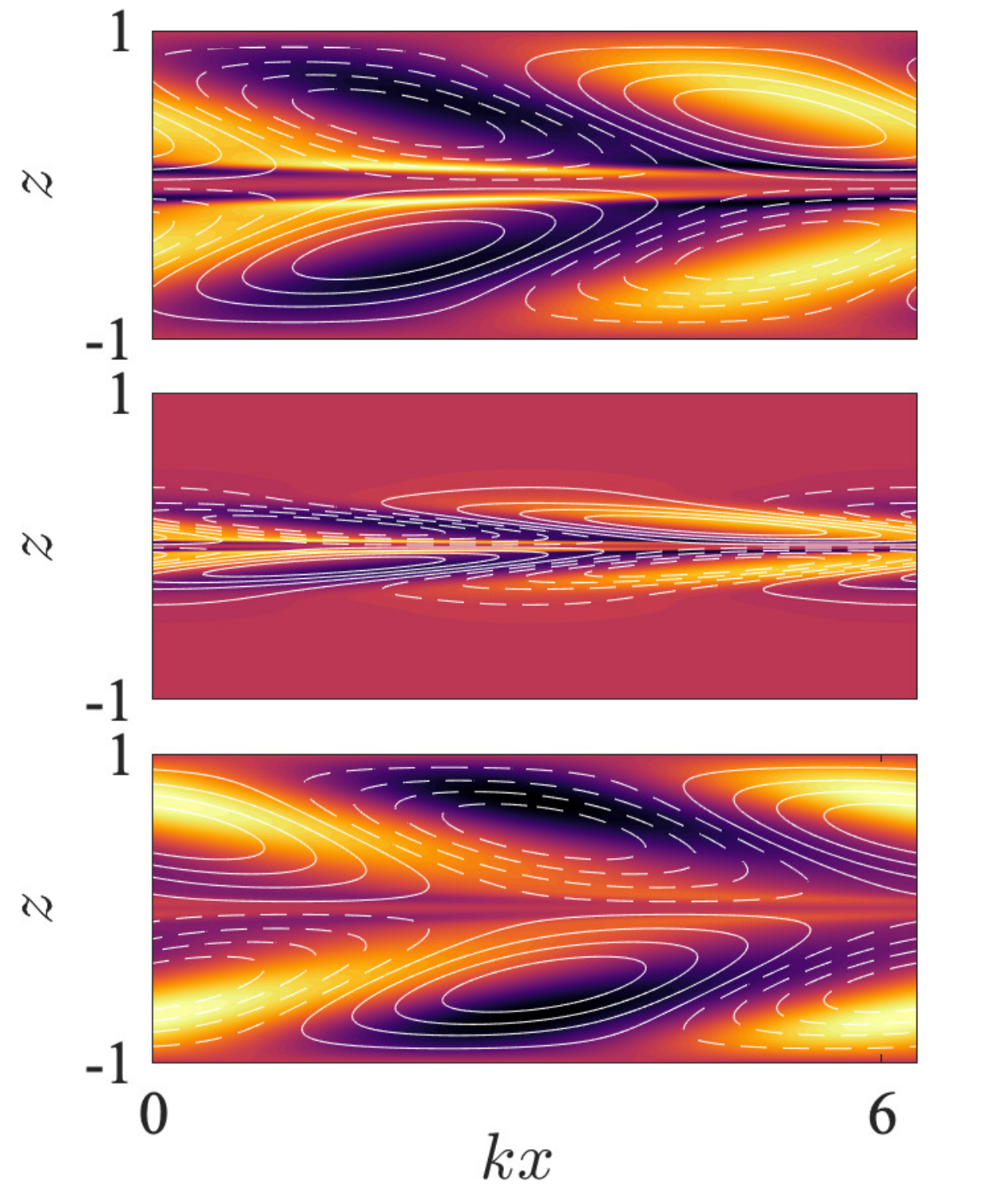}
    \vspace{-5mm}    
    \caption{(Left) Contours of linear growth rate $\sigma$ for most unstable symmetric 
    instability waves in a two-dimensional channel flow of a FENE-P fluid ($L=500$) for streamwise wavenumbers $k\in \mathbb N$.
    The dashed white line indicates marginal stability $\sigma=0$.
    The overlayed grid and symbols identify DNS runs in a box of length $l_x=2\pi$: (blue circle) laminar,
    (red square) EIT, (green triangle) arrowhead.
    The blue lines identify where we have performed arclength continuation of the nonlinear travelling wave born in this instability (at $k=2$). 
    (Right) Visualisation of the instability wave in the $x-z$ plane at points indicated by white circles in the stability diagram. 
    Contours show the trace of the (perturbation) conformation, lines the streamfunction. From top to bottom: $(Re,Wi,k)=(60,30,2)$,
    $(2000,90,7)$ and $(130,90,1)$.}
    \label{fig:stability}
\end{figure*}
%
%
%
Linear stability results for the centre mode instability are reported in figure \ref{fig:stability}. 
These results were obtained by linearising equations (\ref{eqn:governing}) and solving for the complex
frequency $\omega=\omega_r + \zi \sigma$ of normal mode perturbations 
$\boldsymbol \varphi(\mathbf x, t)= \hat{\boldsymbol \varphi}(z)\text{exp}(\zi k x - \zi \omega t)$, where $\boldsymbol \varphi \equiv (u,w,{\rm c}_{xx},{\rm c}_{xz},{\rm c}_{zz},p)$ is a vector of the flow variables.
Since computations are performed in boxes of length $l_x = m\pi$ with $m\in \mathbb N$, we search over integer wavenumbers only. The resulting temporal eigenvalue problem was solved by expanding in $N_c\sim 200$  Chebyshev polynomials over half the channel, $z\in[-1,0]$, and applying  symmetry conditions at $z=0$ ($u$ symmetric, $w$ antisymmetric).

The centre mode first becomes unstable at $(Re,Wi,k) \sim (50,25,2)$; the associated eigenfunction (also shown in figure \ref{fig:stability}) consists of trains of tilted vortices of opposite sign either side of $z=0$. On the upper branch of the $\sigma=0$ curve, the instability moves to increasingly high wavenumbers and becomes localized at the channel centreline (for more on the scalings in pipe flow see \cite{Garg2018}). 

We have conducted a number of complementary DNS calculations in a box of length $l_x=2\pi$ at $Re = 1000$
in which we attempt to trigger EIT by applying suction and blowing at the walls \citep[see][]{Dubief2013,Sid2018};
the results are overlayed on the stability diagram in figure \ref{fig:stability}.
The calculations include a large region of parameter space where the flow is predicted to be linearly stable, 
and EIT is obtained for modest $Wi$ prior to the emergence and stabilisation of a single domain-filling arrowhead structure  (either steady or weakly periodic in time) as
$Wi$ increases. 
In regions of instability, the attractor is always an arrowhead, which would be consistent with a connection to the linear bifurcation. 
However, the $k=1$ mode is stable in all the parameter configurations used in the DNS calculations, clearly implying subcriticality.

%
%
\begin{figure}
    \centering
    \includegraphics[width=0.48\textwidth]{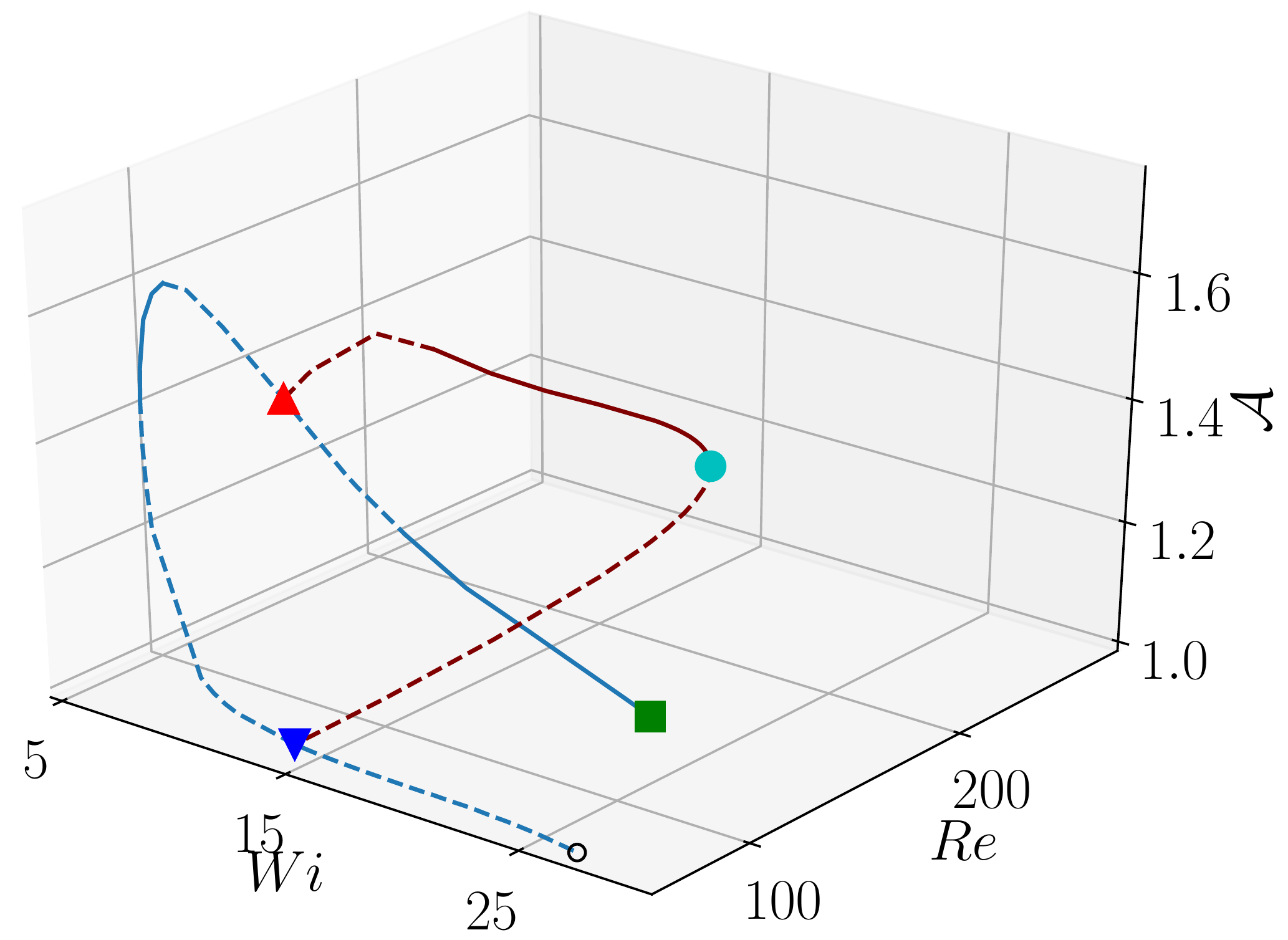}
    \vspace{-5mm}
    \caption{Arclength continuation in $Wi$ at $Re=60$ (blue) and in $Re$ at $Wi=15$ (maroon) in a box of length $l_x=\pi$. The amplitude $\mathcal A:= \langle\text{tr}\mathbf C\rangle_V / \langle \text{tr}\mathbf C_{\text{lam}}\rangle_V$. The solution curve was initialised at $(Re,Wi)=(60,30)$ by perturbing a laminar flow with the linear instability wave at $k=2$, which was allowed to saturate onto the stable upper branch solution. 
    The saddle node at $Re=60$ is at $Wi=8.77$; the saddle node at $Wi=15$ is at $Re=253.71$.
    The linear bifurcation point at $(Re,Wi)=(60,26.9)$ is identified with a small black circle.
    Snapshots of the the TW at points highlighted on the curve are shown in figure \ref{fig:states}.}
    \label{fig:WiRe_cont}
\end{figure}

\begin{figure}
    \centering
    \includegraphics[width=0.235\textwidth]{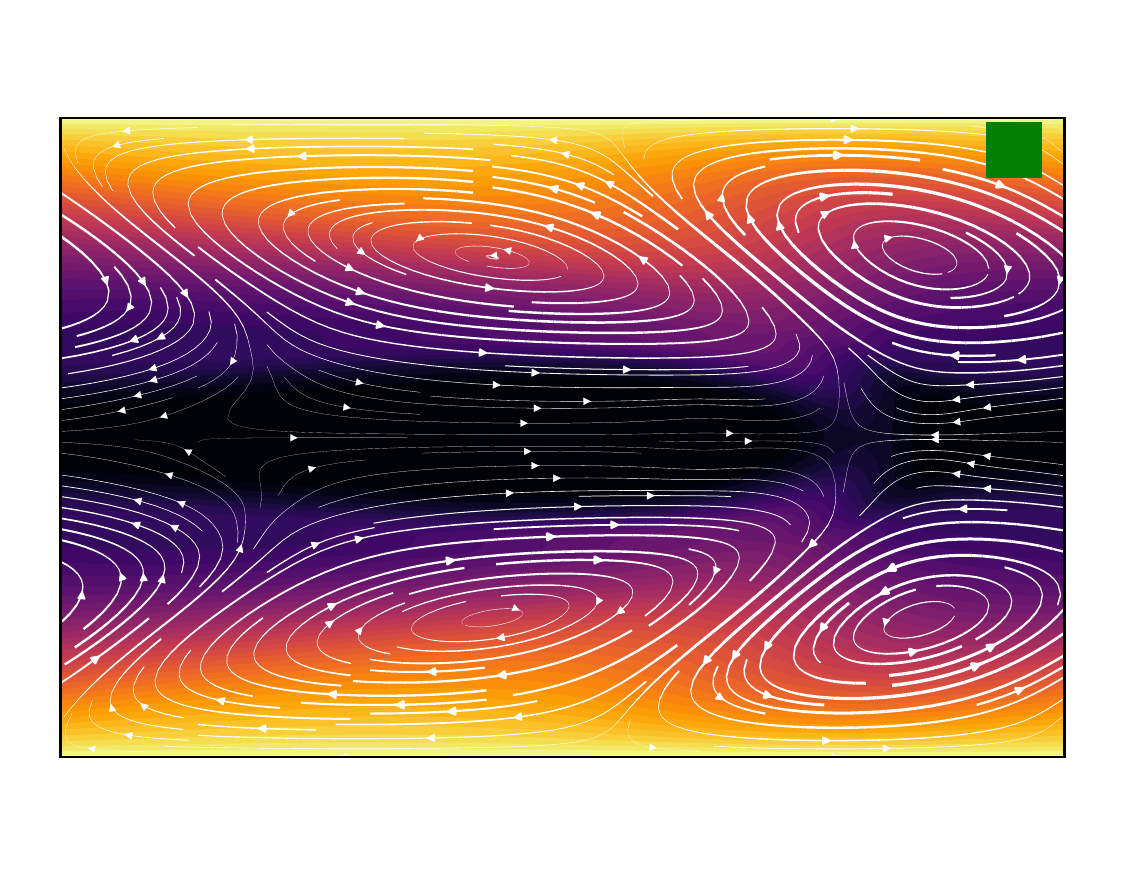}
    \includegraphics[width=0.235\textwidth]{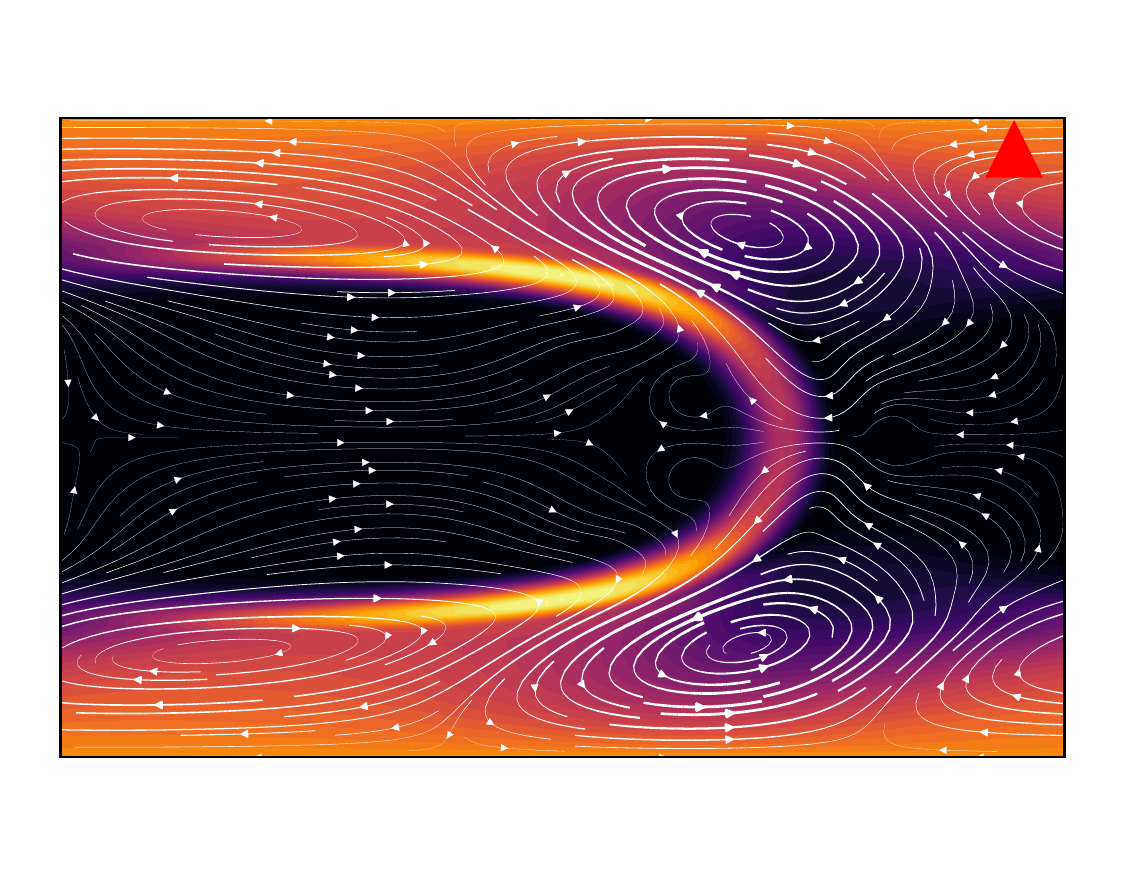}
    \includegraphics[width=0.235\textwidth]{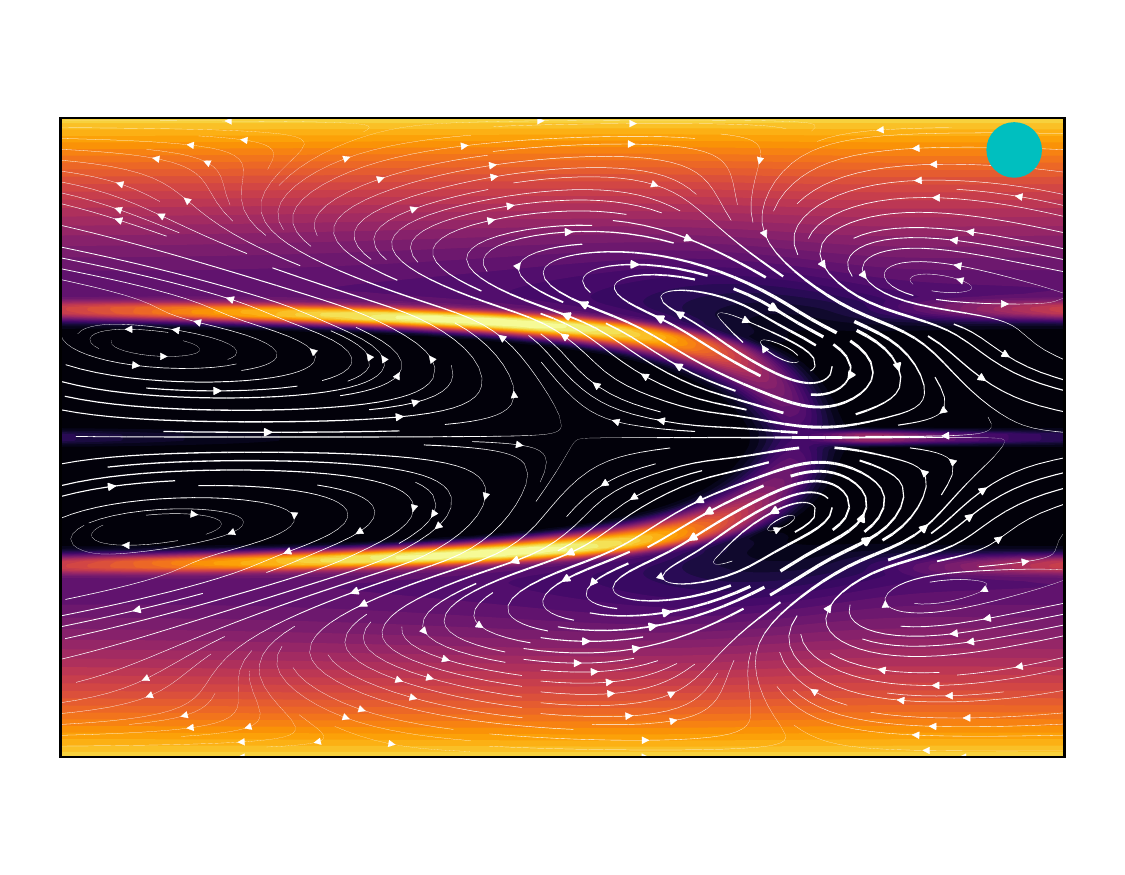}
    \includegraphics[width=0.235\textwidth]{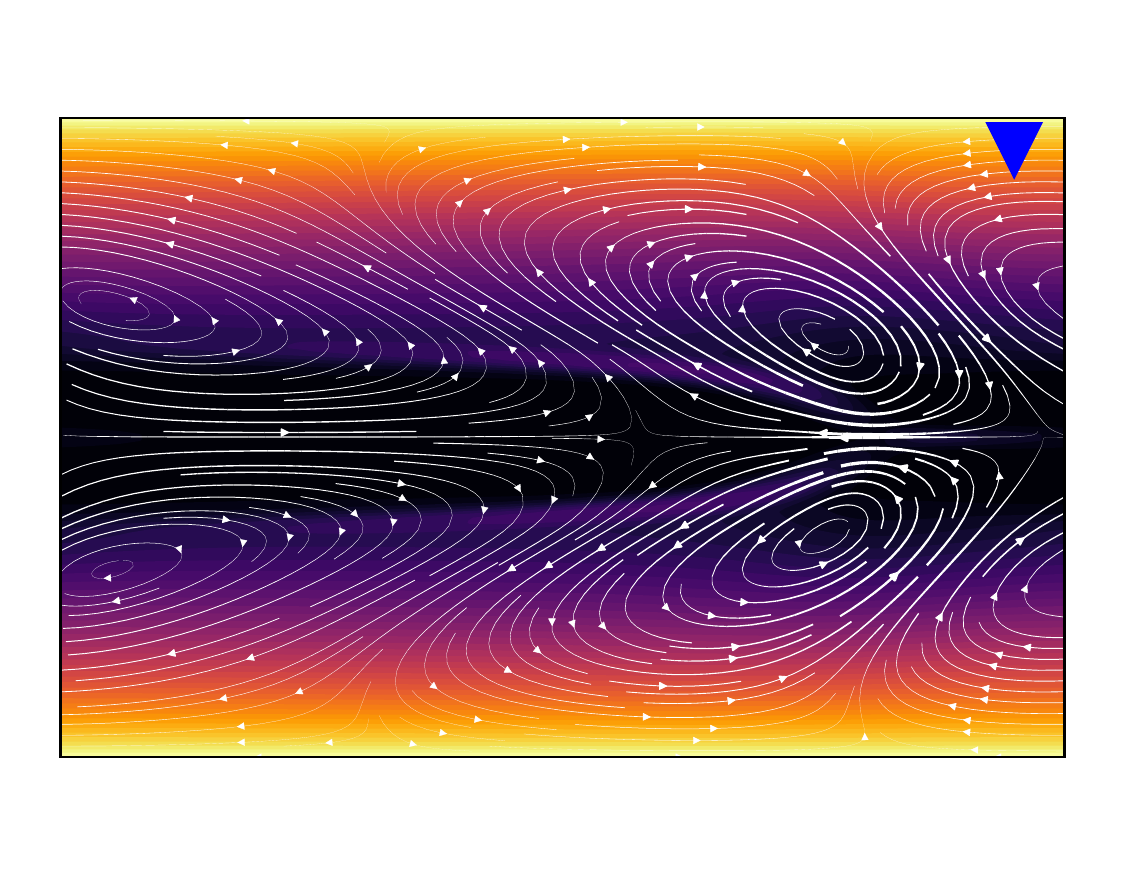}
    \vspace{-5mm}
    \caption{Travelling waves corresponding to points identified in figure \ref{fig:WiRe_cont}. Contours show $\text{tr}\mathbf C/L^2$, lines are the perturbation ($\mathbf u - \mathbf U_{\text{lam}}$) streamfunction. The aspect ratio matches the snapshot of the long domain ($l_x=4\pi$) calculation reported in figure \ref{fig:long}.}
    \label{fig:states}
\end{figure}

To substantiate the connection of the arrowhead solution to the centre mode bifurcation and show the bifurcation's significant subcritical nature, we take the centre mode eigenfunction just beyond the point of marginal stability, $(Re,Wi,k) = (60,30,2)$, and apply it as a perturbation to the laminar flow in a $l_x=\pi$ box. Timestepping leads to saturation onto a stable TW which shares some similarity to the linearised eigenfunction, although with  a conformation field which is significantly perturbed (note the amplitude in figure \ref{fig:WiRe_cont} and the snapshot of the TW in figure \ref{fig:states}). The TW readily converges in a Newton solver looking for a steady solution in a Galilean frame and then can be arclength-continued around in $Wi$ while holding $Re$ fixed: see figure \ref{fig:WiRe_cont}. The amplitude of the travelling wave initially increases as $Wi$ drops with a saddle node bifurcation reached at $Wi\approx 8.8$ and a (very) low amplitude lower branch connects back to the bifurcation point at $Wi\approx 26.9$. The upper branch TW  has a Hopf bifurcation at $Wi\approx 20$ below which the attractor is a simple (relative) periodic orbit with period $T=O(100)$ ($T\sim 250$ at $Wi=15$) and the TW restabilises at $Wi \approx 10$.

The structure of the (unstable) TW at the subcritical pair $(Re,Wi)=(60,15)$ is shown in figure \ref{fig:states} for both the upper and lower branches. On the upper branch, resemblance to the linear stability wave is largely lost and the $\text{tr}\mathbf C$ field has adopted the arrowhead form: a single curved sheet of highly stretched polymer runs across the channel centreline (compare with the structure in the $l_x=4\pi$ domain of figure \ref{fig:long}) and the flow field is mirror symmetric about $z=0$.  The lower branch state does not resemble the arrowhead and is somewhat closer to the linear eigenfunction, although there is a pair of weak  sheets of polymer stretch clearly visible.

To probe the connection of the arrowhead emerging from the centre mode bifurcation to EIT, we also take the TW at $Wi=15$ and continue up in $Re$: see the
maroon curve in figure \ref{fig:WiRe_cont}. The TW is unstable up to $Re\approx 115$ -- whether the periodic orbit exists subcritically beyond this point has not yet been investigated.  At this $Wi$, the saddle node sits at $Re\approx 254$ where, again, the state takes the shape of an arrowhead in polymer stretch: see figure \ref{fig:states}. The sheets either side of the centreline have moved inwards relative to their position at $Re=60$, though this movement is not monotonic with increasing $Re$. 

%
%

Further work is required to establish the self-sustaining mechanism that produces the arrowhead, though the parameter values for which it is observed indicate that elasto-inertial wave propagation along tensioned streamlines may play a role (see the linear mechanisms discussed in \cite{Page2015,Page2016}); this may also help to establish the $z$-locations of the parallel sheets of polymer stretch that make up the  `edges' of the arrowhead. Other recent studies have argued for the importance of structures connected to Newtonian Tollmien-Schlichting (TS) waves \cite{Shekar2019} in EIT, but have not been able to explicitly continue these TWs around in the parameter space. These studies have been performed at much higher Reynolds numbers and higher values of the solvent viscosity than those considered here.  Continuation of both the arrowhead and TS TWs in longer boxes will help establish where these dynamics overlap. 


In summary, we have isolated the first exact coherent structures in viscoelastic channel flow by performing arclength continuation from 
the recently discovered high-$Wi$ instability reported in \cite{Garg2018}.
Our computations have demonstrated that the bifurcation is strongly subcritical in both $Wi$ and $Re$. 
The upper branch solutions take the form of large arrowhead structures in the polymer stress field -- structures which 
are observed intermittently in computations of EIT in large boxes and which have been observed to be stable attractors at very high $Wi$. Beyond indicating that the origins of EIT are purely elastic in nature and so disconnected from Newtonian dynamics, more importantly, these exact coherent structures provide a crucial  beachhead to  identify the self-sustaining processes which underpin EIT and also possible connections to elastic turbulence.


%

\end{document}